# Spin-Transfer Excitations of Permalloy Nanopillars for Large Applied Currents

S. I. Kiselev[a], J. C. Sankey, I. N. Krivorotov, N. C. Emley, A. G. F. Garcia, R. A. Buhrman, and D. C. Ralph

Cornell University, Ithaca, New York, 14853 USA

Using measurements of the spectra of microwave-frequency resistance oscillations, we determine the phase diagram of magnetic excitations caused by torques from DC spin-polarized currents in (thin permalloy)/copper/(thick permalloy) multilayer samples. We extend the measurements to larger values of current than have been reported previously. We find several additional modes that we are able to identify with motion of the thick magnetic layer as well as the thin one. Peaks in the microwave spectra at multiple frequencies suggest that spatially non-uniform dynamical states can be important in some circumstances. We compare the experimental phase diagram with simple theoretical models and achieve a good qualitative agreement.

72.25.Ba, 73.40.-c, 75.40.Gb

## INTRODUCTION

A spin-polarized current passing through a ferromagnetic layer can transfer angular momentum so as to apply a torque directly to the layer's magnetic moment.[1-9] This torque is of interest as a more efficient way of manipulating the moments of small magnets, compared to employing magnetic fields, in possible applications that include nonvolatile magnetic memory and microwave oscillators. Spin-transfer torques and the magnetic excitations they produce in magnetic multilayers have been studied in two primary sample geometries: metallic point contacts to a multilayer[10-13] and fabricated devices in which a magnetic multilayer is formed into a pillar structure of order 100 nm in diameter.[14-28] In the nanopillars, the region of the sample excited by the spin-transfer current can be isolated, not part of an extended magnetic film as in the case of the point contacts. This makes the nanopillar geometry simpler to model and to compare with theoretical predictions. Past work has generally focused on nanopillars with a ferromagnet/normal metal/ferromagnet (FNF) geometry, with one of the magnetic layers thicker than the other (or exchange-biased to an antiferromagnetic layer) so that it is relatively insensitive to the spin-transfer torque, while the thinner "free" layer can be reoriented by the torque. In our previous measurements,[19,24] the thick layer was not etched through, so that it remained strongly exchange-coupled to an extended thin film, making it even more resistant to being manipulated by spin transfer. The dynamical phase diagram for such devices has been established for low-to-moderate currents (up to 4 times the critical current to first produce spin-transfer-driven dynamics in the thin

---

[a] current address: Hitachi Global Storage Technologies, San Jose, CA, 95120, USA

magnetic layer) through a combination of quasi-static and microwave-frequency resistance measurements.[19,24] Depending on the applied current and the magnetic field, the spin-transfer torque can act either to switch the thin magnetic layer between static states or to excite the thin layer into a variety of distinguishable steady-state dynamical modes.

In this article we use microwave-frequency measurements of resistance oscillations to extend to higher currents ($I$) the characterization of the dynamical modes that can be excited in F/N/F nanopillar devices by spin-transfer torques. We reach a higher-current regime by employing permalloy (Py=$Ni_{80}Fe_{20}$) magnetic layers, whereas in our previous studies with in-plane magnetic fields we used Co.[19] The smaller magnetic moment and weaker magnetic anisotropy in Py as compared to Co leads to smaller critical currents $I_c$ for the excitation of dynamic states in Py. This allows us to study a much larger range of $I/I_c$ for the range of DC bias currents that can be applied safely to the sample before electromigration-induced failures become a danger (< 10 mA). We also use devices in which the thick magnetic layer is etched through, which makes it easier to excite precessional dynamics of the thick layer as well as the thin layer, at accessible current levels. Our primary new results are: (a) The permalloy samples exhibit a spin-transfer-induced dynamical mode for the thinner layer that we did not observe previously in cobalt samples. (b) At sufficiently large currents, we observe dynamical modes involving motion of the thicker "fixed"-layer magnetic moment as well as the thin-layer moment. These modes are distinguished by very narrow peaks in the frequency spectrum of resistance oscillations (FWHM~10 MHz) compared to the lower-current thin-layer modes, and also frequencies that are distinct from the thin-layer modes. (c) When the thick layer is excited, the microwave spectrum often exhibits peaks at multiple frequencies that are not related as harmonics, suggesting that the dynamics include not just spatially uniform precessional states, but also non-uniform spin-wave modes. (d) We compare our measurements to dynamical phase diagrams calculated numerically by assuming that the Slonczewski model of spin torques[1] can be applied to both of the magnetic layers in our devices, and we find good qualitative agreement over most of the phase diagram.

## SAMPLE FABRICATION AND LOW-FREQUENCY CHARACTERIZATION

Our samples are made by sputtering a multilayer of 80 nm Cu/20 nm Py/6 nm Cu/2 nm Py/2 nm Cu/30 nm Pt (Py=$Ni_{80}Fe_{20}$), and then using electron-beam lithography and ion milling to pattern a pillar-shaped structure with an approximately elliptical cross section of 130 nm × 70 nm. We mill completely through both Py layers so that neither is coupled to an extended magnetic film. A portion of the Cu underlayer is left unetched so that it may serve as the bottom electrode. We then planarize with silicon oxide, and deposit Cu to make a top contact to the Pt cap of the nanopillar.[29] All transport measurements are taken with current perpendicular to the layers of the nanopillar. By convention, positive current corresponds to electron flow from the thinner to the thicker Py layer. While we will focus in this paper on a single sample, we observed very similar behavior in two other samples that we studied in detail.

We perform an initial characterization of the samples by standard measurements of the differential resistance as a function of $I$ and of the applied magnetic field, $H$ (Fig. 1 and 2). The results are qualitatively very similar to previous low-current studies of similar devices. Consider first the case of $H$ applied in the plane of the sample layers, along the magnetic easy axis, for which the room-temperature magnetoresistance near $I$=0 is plotted in Fig. 1(a). As $H$ is decreased from large positive fields (dark curve), the increase in resistance near $H$ = 0.38 kOe corresponds to the switching of the thin Py layer from parallel (P) to antiparallel (AP) alignment with respect to the thick-layer moment. The total magnetic moment of the Py thin layer is sufficiently small that this layer is superparamagnetic, so that this transition is reversible. The switching field denotes the condition that the sum of the dipole field, $H_d$ on the thin layer due to the magnetization of the thick film plus the applied field is equal to zero. We can therefore identify $H_d$ = -0.38 kOe. As $H$ is decreased further, starting from positive values, the thick layer switches at $H$ = -0.85 kOe to return the sample to the P configuration. This field is close to the value we estimate for the shape anisotropy of the thick layer, 0.77 kOe (see below), with a small correction due to the dipole field of the thin layer acting on the thick layer.

Figure 1(b) shows the room-temperature differential resistance of the sample as a function of $I$ and in-plane $H$ applied along the easy axis, when the sample is prepared with the thick-layer moment aligned in the field direction. Because the Py thin layer is superparamagnetic, we do not observe the hysteretic switching as a function of $I$ that was measured previously for samples with Co thin layers and $H$ less than the coercive field. Instead, near $H$=0 the sample simply switches reversibly between resistance values corresponding to P and AP alignment. For larger $H$ ($H$ > 0.4 kOe), as a function of $I$ the sample resistance changes at a critical current $I_c(H)$ from the P value to a value that is intermediate between the P and AP resistances. This is very similar to the behavior of devices with Co thin layers, for which we identified this transition as due to the onset of current-driven precession of the thin layer.[19] In many samples this transition is associated with a single peak in $dV/dI$ as a function of $I$; but for the sample on which we focus here, multiple peaks in $dV/dI$ can be seen for $H$ > 7 kOe.

By applying $H$ perpendicular to the sample plane we can gradually change the relative angle of the magnetic moments in the two layers from the AP configuration with moments in the sample plane near $H$=0 (due to dipolar coupling between the layers) to the P configuration with moments perpendicular to the sample plane at $H$ > 5.2 kOe (Fig. 2(a) inset). The differential resistance for $H$ perpendicular to the sample plane is shown for selected values of $H$ in Fig. 2(a) and as a function of $H$ and $I$ in Fig. 2(b). These data share many features with the data we have presented previously for a sample with a Py thin layer and a Co thick layer.[24] At low $H$ (< 4.5 kOe), as a function of $I$, there are changes in $dV/dI$ that we have identified as due to the onset of small angle precession, followed by a large peak in $dV/dI$ due to a transition in which the thin-layer moment evolves from a precessional state to a static state approximately anti-aligned with $H$. For $H$ > 4.5 kOe the peak in $dV/dI$ corresponds to a simple transition of the thin layer moment between two static states P and AP with respect to the magnetic field. A class of features, that were not seen in our previous work on samples with a Co thick layer, are a set of lines at large positive current at which $dV/dI$ exhibits dips. These lines of dips can first be observed beyond a line of critical currents we will call $I_{c2}(H)$, and they shift to lower $H$ with increasing $I$. The onset line of critical currents $I_{c2}(H)$ increases approximately linearly as a function of *H*. The dotted line in Fig. 2(b) shows the position of $I_{c2}(H)$, as determined by the appearance of a first small dip in *dV/dI* (too small to be clearly visible in Fig. 2(b)) that is coincident with the onset of a sharp microwave mode discussed below. Sometimes similar dips in *dV/dI* are also observed at negative currents in the same field range, but their amplitude varies between samples and is about an order of magnitude smaller than the larger dips in *dV/dI* at positive currents. All of these features associated with the dips in *dV/dI* were reported previously by the NYU/IBM collaboration, who argued that they were due to thick-layer excitations.[28] This conclusion is supported by recent calculations.[30] Below, we will present high-frequency measurements which provide direct evidence that this interpretation is correct.

## PROCEDURES FOR MICROWAVE MEASUREMENTS

In order to determine the types of magnetic dynamics excited by spin-transfer torques in our samples, we perform frequency-domain measurements of microwave signals generated by the magnetization dynamics.[19] We couple the sample to a semi-rigid 50 Ω coaxial line by ribbon wire bonds, and apply a DC current bias through a bias-T. The spin-transfer torques can cause the magnetic moments in the sample to reorient, producing changes in the sample resistance as a function of time on microwave-frequency scales. Because of the DC current bias, these resistance changes generate microwave-frequency voltage signals. We detect these signals by amplifying them with 40 dB amplifiers operating in either the 0.5-18 GHz or 18-40 GHz range and then frequency-analyzing with either a commercial spectrum analyzer or a homemade heterodyne circuit.[19] The homemade circuit allows us to increase the resolution bandwidth of the measurement beyond the limit allowed by our commercial spectrum analyzer, thereby increasing the signal-to-noise ratio when we measure spectra with broad peaks. We are careful to make the resolution bandwidth of our circuit sufficiently narrow that it does not produce an artificial broadening of the signals under measurement. Before plotting any of the measured spectra, we subtract the noise background measured at *I*=0 (due to Johnson noise and amplifier noise) and we correct for the frequency-dependent gain of the spectrometer. We calibrate the frequency-dependent gain by measuring the Johnson noise from a 50 Ω resistor at room temperature and liquid-nitrogen temperature, as well as the spectrometer noise when the circuit is connected to a short circuit and an open circuit.[19] We divide most of the spectra shown below by $I^2$ to facilitate comparisons of the magnitude of the resistance oscillations at different *I*.

## MAGNETIC DYNAMICS FOR IN-PLANE MAGNETIC FIELDS

We find that the spin transfer torque can drive the magnetic layers in our samples into several distinct types of dynamical modes, depending on the magnitudes of *I* and *H*. In this section, we will focus on in-plane magnetic fields and illustrate the dynamics that can occur at the selected field values *H* = 0.7, 1.3, and 6 kOe. We will emphasize the ways in which our Py/Cu/Py samples differ from the Co/Cu/Co nanopillar samples we examined previously.[19]

**$H = 0.7$ kOe.**

Figure 3(a) displays the microwave power density measured for several values of $I$ at $H = 0.7$ kOe. At $I = I_c = 1.4$ mA we begin to resolve broad peaks in the spectrum with a fundamental frequency $f = 3.5$ GHz. The peaks shift to higher frequency as a function of magnetic field, consistent with expectations for precession of the thin layer. For $I < 7$ mA, the peaks in Fig. 3(a) ride on top of a large low-frequency tail, which in some samples completely dominates the power spectrum. We believe that this tail can be explained by stochastic switching between different precessional states, or between a precessional state and a static state.[23] The behaviors of both the thin layer precessional peaks and the low-frequency background are qualitatively very similar in Py/Cu/Py and Co/Cu/Co devices.[19]

As a function of increasing $I$ up to 2.6 mA, the frequency of the precessional peak decreases. This behavior was also seen in Co/Cu/Co devices, and by comparison with dynamical simulations it was identified with large-angle "clamshell" precession of the thin layer.[19] In Co/Cu/Co, the frequency shifted toward zero as $I$ increased, after which point the microwave signals suddenly ceased, even though simulations suggested that spin-transfer should next excite a single-domain magnet into an out-of-plane approximately circular orbit with a large microwave signal. We speculated that this difference was due to the breakdown of single-domain behavior in the Co thin layer.[19] In the Py/Cu/Py samples, we observe something different. For $I$ greater than 2.6 mA, the frequency of the fundamental and its harmonics begin to increase (e.g., for $I = 3.6$ mA and 5.4 mA shown in Fig. 3(a)), and then eventually decrease again at larger bias (e.g. $I = 6.4$ mA). The peaks in the microwave spectrum in this regime become larger and narrower than at lower currents, corresponding to a larger-amplitude more coherent precessional motion. At its narrowest, the fundamental peak in this sample has a full width at half-maximum (FWHM) of ~700 MHz.[31]

For even larger $I$, at approximately 6.8 mA, the amplitudes of the microwave signals decrease rapidly to zero, with both the low-frequency tail and the peaks that we have associated with the thin-layer dynamics disappearing simultaneously. However, at and beyond this current we observe two new peaks, much narrower than the lower-current signals. These new peaks are marked by dots in Fig. 3(a). They appear initially at 9.26 GHz (we will call this peak A) and 12.56 GHz (peak B). Because these frequencies are not related as harmonics, we interpret them as due to separate modes. The disappearance of the broad low-frequency signals and the appearance of the two narrow new peaks occurs simultaneously with the transition shown in Fig. 1(b) to the approximately AP resistance state.

Detailed views of the spectral signals for peaks A and B at $H = 0.7$ kOe and $I = 7$ mA are shown as insets to Fig. 3(b). The lower-frequency peak A has a FWHM of 11 MHz and peak B has a FWHM of 59 MHz. Both peaks can be fitted accurately with a Lorentzian peak shape. Figures 3(c) and 3(d) display the evolution of the power density in the peaks as a function of $I$. The amplitude of the lower frequency mode (mode A) grows continuously from 0.04 pW/mA$^2$GHz at $I = 6.8$ mA to 1.7 pW/mA$^2$GHz at $I = 10$ mA, while its FWHM decreases from 56 MHz to 10 MHz and frequency decreases slightly from 9.26 GHz to 9.19 GHz. The power density in the higher frequency mode (mode B) grows quickly after onset until it reaches a maximum of 0.11 pW/mA$^2$GHz (with FWHM = 64 MHz) at $I = 7.5$ mA, then it decreases slowly to 0.05 pW/mA$^2$GHz (with FWHM = 59 MHz) at $I = 10$ mA, while the peak frequency also decreases slowly from 12.56 GHz at 6.8 mA to 12.47 GHz at 10 mA. The magnetic field dependence of the frequencies of peaks A and B are plotted in Fig. 3(b), measured at the current for which the peaks are first visible. The threshold current needed to excite the sharp peaks grows quickly as a function of $H$ and becomes greater than $I = 10$ mA at $H = 1.2$ kOe, so we did not attempt to follow the evolution of the peaks beyond this field value. The difference in frequencies between modes A and B is about 3.3 GHz and varies little as a function of magnetic field or current.

*H = 1.3 kOe.*

Beyond $H = 1.2$ kOe we observe only the broader, low-frequency microwave signals in the range of current (I ≤ 10 mA) that we have explored. Figure 3(e) shows the microwave power density as a function of $I$ and $f$. At the onset of the dynamical signal near $I = 1.6$ mA in Fig. 3(e), the measured signal consists of a broad peak centered at 8.6 GHz and a low-frequency tail. As $I$ is increased, the frequency of the precessional peak decreases and eventually it is not distinguishable from the low-frequency background. In simulations of the magnetic dynamics, this decrease in frequency is characteristic of the large-angle "clamshell" orbit.[19] Then for $I$ greater than 4 mA, Fig. 3(e) shows the emergence of a broad peak the frequency of which increases with current, while the amplitude of the low-frequency background is decreased. This regime of precession with a frequency that increases with I was not observed previously in Co/Cu/Co samples.[19] In this new mode the microwave power reaches a maximum at $I = 7.2$ mA and then decreases slowly. As we increase $H$, the maximum microwave power in the new state is reduced (1.7 pW/mA$^2$ at $H = 0.7$ kOe vs. 0.8 pW/mA$^2$ at 1.3 kOe). We will analyze the physical nature of this mode in the Discussion section below.

*H = 6 kOe*

For $H > 4.5$ kOe we do not observe any precessional dynamics that we can associate with the thin layer. At the critical current where the sample undergoes a transition from the P to the AP resistance state (Fig. 1(b)), we find only a small stochastic-noise low-frequency tail in the microwave spectrum. However, at significantly larger currents, ($I > 8.5$ mA, for the fields 5.5 kOe < $H$ < 6.5 kOe) we again detect a narrow microwave mode, whose width and a frequency depend on $I$ in a manner similar to mode A described above. As a function of magnetic field, the frequency of this mode appears to fall along the same curve as the signals for mode A observed for H < 1.2 kOe (Fig. 3(b)). At $H = 6$ kOe the narrow mode first appears at $I = 8.5$ mA with frequency 25.997 GHz. Its amplitude grows with increasing $I$ and reaches 0.0044 pW/mA$^2$GHz at $I$=10 mA, while its frequency decreases to 25.992 GHz (FWHM~ 7 MHz). This maximum intensity is two orders of magnitude smaller than the maximum magnitude of peak A at 0.7 kOe.

**Discussion**

We can identify the narrow peaks in the microwave spectra observed at large $I$ as involving the dynamics of the thick layer, based on (1) the fact that they appear for positive currents when the thin-layer moment is approximately AP to the thick-layer moment, (2) the dependence of the frequency of the modes on *H*, and (3), the narrow widths of the peaks.

When the two magnetic layers in a nanopillar device have moments that are approximately parallel and a positive current is applied, the sign of the spin-transfer torque on the thin layer is such that this torque can turn the thin-layer moment away from the direction of the thick-layer moment or excite spatially non-uniform motion of magnetization within the thin layer. This is the origin of the microwave signals that we observe at low currents. However, when the current has a sufficiently-large value that the thin-layer moment is reoriented approximately antiparallel to the thick-layer moment, then positive currents apply a torque on the thin-layer so as to stabilize the uniform antiparallel orientation. In this situation, positive current also applies a torque on the thick-layer moment so as to push it away from the thin-layer moment or to excite non-uniform excitations within the thick layer. Therefore, for large positive currents applied to a sample in which the magnetic layers are antiparallel, spin-transfer theory predicts that it is the torque on the thick layer that should drive any instabilities toward dynamical states.[1-3] Because the two layers remain strongly coupled by their mutual spin-transfer torques, the resulting dynamics should involve the coordinated motion of the magnetic moments in both layers.

Let us now consider the frequency of mode A as a function of magnetic field. The Kittel expression for the frequency of small angle precession for a magnetic film is:[32]

$$f^2 = (g\mu_B/2\pi\hbar)^2 (H + H_d + H_{an})(H + H_d + H_{an} + 4\pi M_{eff}), \tag{1}$$

where $\mu_B$ is the Bohr magneton, $H_{an}$ is the in-plane uniaxial anisotropy of the precessing layer, $4\pi M_{eff}$ is effective demagnetization field of the layer, and $H_d$ is the dipole field acting on the layer. In considering precession of the thick layer, we will assume that $H_d$ is small because of the small magnetic moment of the thin layer. Assuming for simplicity that g = 2.0, we can fit the frequency of mode A simultaneously in the field ranges $H < 1.2$ kOe and 5.5 kOe $< H < 6.5$ kOe using the same fit parameters: $4\pi M_{eff} = 3.9 \pm 0.8$ kOe and $H_{an} = 1.3 \pm 0.3$ kOe (black curve in Fig. 3(b)). These values can be checked against the parameters that are expected for the thick layer, if we assume that shape anisotropy is dominant; in terms of the demagnetization factors $N_x$, $N_y$, and $N_z$ (defined using the convention that $N_x + N_y + N_z = 4\pi$) we can expect $4\pi M_{eff} \approx (N_z - N_y)M_s$ and $H_{an} \approx (N_y - N_x)M_s$, where $\hat{z}$ is the out-of-plane direction, $\hat{x}$ is the easy-axis direction, and $M_s$ is the saturation magnetization. Using SQUID magnetometry on test layers exposed to the same processing conditions as the measured samples, we measure $M_s = 680$ emu/cm$^3$ ($4\pi M_s = 8.5$ kOe) for 20 nm Py layers. This is reduced from the bulk value 800 emu/cm$^3$ ($4\pi M_s = 10$ kOe), possibly due to diffusion of Cu into the Py during a 170 °C e-beam resist bake. If we approximate the shape of the thick layer magnet by an ellipsoid with axes of 130 nm, 70 nm, and 20 nm, we calculate using the expressions given in ref. [33] that $N_z = 4\pi(0.67)$, $N_y = 4\pi(0.19)$, and $N_x = 4\pi(0.14)$, from which we obtain the estimates $4\pi M_{eff} \approx 4.1$ kOe and $H_{an} \approx 0.43$ kOe for the thick layer. We conclude that the frequencies measured for peak A are in reasonable accord with the expected value of $4\pi M_{eff}$ for the thick-layer moment. The difference between the measured and estimated values of $H_{an}$ may be a consequence of the fact that the true layer shape, an elliptical pancake, is not well approximated by the three-dimensional ellipsoid assumed in the estimate.

Regarding the width in frequency of mode A, it has been observed that the widths of the microwave peaks driven by spin-transfer are affected by thermal fluctuations, becoming broader as the temperature is increased.[34,35] We therefore expect thicker magnetic layers to exhibit narrower peaks, because their dynamics should be influenced less by thermal effects. The width of mode A is 80 times narrower than the peaks associated with thin-layer dynamics observed at lower currents in the same sample. It is therefore natural to associate the narrower peaks with the thick layer.

We find that we cannot explain the measured frequency of mode B by simple small-angle precession of either magnetic layer. However, recent calculations show, for a sample geometry similar to our experiment, that spin-transfer-driven excitation of non-uniform spin-waves may accompany or even precede the excitation of the uniform precessional mode.[30,36,37] An approximate expression for the frequencies of the non-uniform spin wave modes in a ferromagnetic film is given by Herring-Kittel formula:[38]

$$f^2 = (g\mu_B/2\pi\hbar)^2 (H + H_{sw} + H_{an})(H + H_{sw} + H_{an} + 4\pi M_{eff}), \quad (2)$$

with $H_{sw} = (2A/M_s)(2\pi/\lambda)^2$, where $A = 1.3 \times 10^{-6}$ erg/cm is the exchange constant for Py, we use $M_s = 680$ emu/cm$^3$, and $\lambda$ is the spin-wave wavelength. The 3.3 GHz difference between modes A and B can be accounted for using $H_{sw} = 0.95$ kOe. This yields a magnon wavelength of 120 nm, close to the length of long axis of our sample cross section. The fact that mode B is observed only at low magnetic fields ($H < 1.2$ kOe) may indicate that a larger field favors more spatially-uniform dynamical states.

**Phase Diagram for In-Plane Magnetic Field**

To summarize these results, in Fig. 4(a) we construct a phase diagram of the dynamical modes that we observe for a magnetic field applied in the sample plane along the easy axis. The regions labeled P and AP denote where the sample resistance corresponds to the parallel and antiparallel states. In the shaded regions, we observe precessional microwave signals which decrease in frequency with increasing |I|. This is the type of signal we have identified with large-angle "clamshell" precession.[19] The region D corresponds to the precessional state of the thin layer in which the frequency increases as a function of I. This is a state that was not observed in our previous studies of Co/Cu/Co samples. The broad microwave peaks that we associate with thin-layer dynamics turn off at the currents marked by open squares, after which we observe the much narrower peaks that we have argued to be associated with precession of the thick layer, likely coupled to the thin layer. We label with the letter T the regions where the narrow peaks are observed. For $H > 4.5$ kOe, as a function of increasing current we observe a switch of the thin layer from the parallel state to a static high resistance (SHR) state (with resistance close to the AP value) with no visible dynamics in the microwave spectrum. Narrow thick-layer microwave peaks exist in the area surrounded by the open squares between 5.5 kOe and 6.5 kOe. In all the samples that we have studied at high currents, we observed the narrow thick-layer microwave peaks at positive I only after the thin-layer moment was first deflected to be approximately antiparallel to the thick-layer moment. In some samples, instead of clear separation of the regions where dynamics of the thick and thin layers are observed, we see a coexistence of narrow thick-layer peaks with broader signals that have frequencies corresponding to thin-layer precession about an axis approximately antiparallel to the thick-layer moment.

# MAGNETIC DYNAMICS FOR OUT-OF-PLANE MAGNETIC FIELDS

For a magnetic field applied perpendicular to the sample plane, it is known from previous studies that there are two qualitatively different regimes when considering the thin-layer dynamics driven by spin transfer.[18,24] At low fields, as function of increasing current, spin transfer can drive a transition from the low resistance (approximately parallel) state, through a series of precessional states for the thin layer, to a static high resistance (SHR, approximately antiparallel) configuration. At sufficiently large magnetic fields, no stable precessional state exists for the thin layer, and one measures a direct transition of the thin layer from a static low-resistance state (SLR, thin layer magnetization parallel to $H$) to a SHR state (thin layer antiparallel to $H$). By going to larger values of current bias, we show with the data presented below that spin transfer can excite narrow thick-layer peaks in the microwave spectrum in both the low-field and high-field regimes. For smaller magnetic fields, the thin and thick layers can precess simultaneously, while at larger field, the thick layer dynamics are observed only well after the thin-layer moment flips from P to AP alignment with respect to the thick layer.

## $H$ = 3 kOe

We will start with discussion of microwave dynamics at $H$ = 3 kOe (Fig. 5(a)), which are characteristic of the lower-field regime. At low currents, the free layer progresses through the same sequence of states determined in a previous study.[24] As $I$ is swept from zero to positive values, the dynamics of the thin layer moment begin at $I$ = 0.8 mA. They evolve from small-angle precession near the $I$=0 equilibrium orientation to precession about the direction of the magnetic field. This transition is marked by a frequency step (from 2.1 to 5.2 GHz) coincident with a dip in $dV/dI$ at $I$ = 1.6 mA. As $I$ is increased beyond 1.6 mA, the magnitude of the resistance oscillations grows, the frequency of resonant peak increases, and its width narrows. Near $I$ = 3.4 mA, there is a jump in frequency from 7 to about 18 GHz, caused by a change in sign of the demagnetization field as the thin-layer moment changes from precessing nearly parallel to $H$ to precessing nearly antiparallel to $H$. This reorientation also produces a sharp peak in $dV/dI$. At larger currents, the amplitude of the thin-layer resistance oscillations decreases until they are no longer visible beyond $I$ = 6.2 mA. Power spectra representative of these various dynamical regimes are shown in Fig. 5(b).

As soon as the thin layer makes its transition at $I$ = 3.4 mA to an alignment antiparallel to the magnetic field (and approximately antiparallel to the thick layer), two narrow microwave modes appear in Fig. 5(a) at 8.7 GHz and 9.3 GHz. The FWHM of each begins at 100 MHz and decreases rapidly with increasing $I$, to a value of 20 MHz at $I$ = 10 mA. We identify these narrow peaks with thick-layer oscillations for the same reasons as in the case of in-plane magnetic field: they occur for conditions under which the spin-transfer torque should destabilize the thick layer (antiparallel configuration and positive $I$), they differ in frequency from the thin-layer modes, and these peaks have widths less than the values typical for thin-layer modes. In the range 3.4 mA < $I$ < 6.2 mA, thin-layer and thick-layer precessional signals are present simultaneously. At even larger $I$, more complicated dynamical behavior of the thick layer is typically observed, including the excitation of additional modes (*e.g.*, see Fig. 5(a) at I > 7 mA).

It is interesting to note that a narrow peak in the microwave spectrum is also observed at large magnitudes of negative current in Fig. 5(a), at a frequency very similar to the thick-layer modes when they are first observed at large positive currents. In this regime the moments of the two magnetic layers are parallel, so that negative currents apply a torque with the sign so as to destabilize the orientation of the thick-layer moment. It is therefore reasonable that thick-layer excitations are seen at negative currents, as well.[28]

*H* = 7 kOe.

In the higher-magnetic-field regime, *H* > 6 kOe, as a function of increasing *I* the thin-layer moment flips directly from a static low-resistance state (approximately parallel to the thick-layer moment) to a high-resistance static state (approximately AP) without the generation of any thin-layer precessional peak in the microwave signal. This is in accord with previous results.[18,24] However, precessional dynamics are still observed for the thick layer at sufficiently large currents. In general, the thick-layer dynamics do not start as soon as the thin-layer undergoes reversal to the approximately AP state; rather, they require a somewhat larger critical current. A typical case, *H* = 7 kOe, is shown in Fig. 5(c). Here the thin layer switches from the low-resistance static state to the high-resistance static state at *I* = 2.4 mA, producing a peak in *dV/dI*. Then, at *I* = 3 mA, two sharp microwave peaks appear at *f* = 7.7 GHz and *f* = 9.4 GHz. The lower-frequency mode (which we shall call mode A1) has about factor of 10 larger amplitude than the higher-frequency mode (B1). At *I* = 4.2 mA, where we observe a small dip in *dV/dI*, the frequencies of both modes undergo an upward kink, and the FWHM of mode A1 doubles from 38 MHz to 55 MHz. At *I* = 5.8, where there is a larger dip in *dV/dI*, mode B1 disappears and mode A1 undergoes a sharp step upward in frequency. Again at *I* = 8.5 mA, there is another large dip in *dV/dI*, and a second upward step in the frequency of mode A1. An analogous behavior has been observed previously for the evolution of the frequency of thin-layer precession in a perpendicular magnetic field, but in that case the kinks and jumps in frequency were associated with peaks, not dips, in *dV/dI*.[24]

**Analysis of the precessional frequencies**

An analysis of the dependence of the precession frequencies on the magnitude of the out-of-plane magnetic field (Fig. 5(d)) confirms our identifications of the different modes. For $H > 4\pi M_{eff}$ for the thick layer, we can fit the *f* vs. *H* dependence for mode A1 with the Kittel expression for uniform small-angle precession of a magnetic film about a magnetic field perpendicular to the sample plane:[32]

$$f^2 = (g\mu_B/2\pi\hbar)^2 (H - 4\pi M_{eff})(H - 4\pi M_{eff} - H_{an}). \tag{3}$$

Here we neglect the dipolar field from the thin layer. Fitting to the data for *H* > 7 kOe using *g* = 2.0, we obtain a good fit with the parameters $4\pi M_{eff} + H_{an}$ = 5.2 ± 0.5 kOe and $4\pi M_{eff}$ = 3.1 ± 0.3 kOe. These are in reasonable agreement with the values $H_{an} + 4\pi M_{eff}$ = 5.2 ± 0.9 kOe and $4\pi M_{eff}$ = 3.9 ± 0.8 kOe obtained from the fit to the field-in-plane data for the thick-layer precession. This agreement allows us to identify mode A1 as uniform precession of the thick layer, likely coupled strongly to the thin layer. The second narrow microwave peak at higher frequency, mode B1, is likely due to a non-uniform precessional mode of the thick layer, in analogy to mode B in the field-in-plane case.

Figure 5(d) also shows (with open triangles) the maximum frequency at which precession of the thin layer is observed at large $I$, after spin-transfer has forced the thin layer to precess approximately antiparallel to the magnetic field. These data were not accessible in our previous studies of Py/Cu/Co nanopillars with an un-etched thick Co layer, because the microwave signal of the thin layer for that sample consisted of multiple peaks rather than a clear precessional signal.[24] The line through the open triangles in Fig. 5(d) is a fit to simplified Kittel expression for a thin-film magnet with $H$ perpendicular to the film, when the film's magnetic moment is AP to $H$:

$$f = g\mu_B(H + 4\pi M_{thin})/(2\pi\hbar), \quad (4)$$

where $4\pi M_{thin}$ is effective demagnetization field of the thin layer. The fit gives $g = 2.25 \pm 0.12$ and $4\pi M_{thin} = 4.5 \pm 0.5$ kOe. The value of the effective demagnetization field is similar to the value $4\pi M_{thin} = 5.1 \pm 0.1$ kOe that we determined previously for a slightly thicker (3 nm) Py thin layer.[24]

**Phase Diagram for Out-of-Plane Magnetic Field**

In Fig. 6(a) we plot the experimental phase diagram for the dynamical modes in perpendicular magnetic field. At positive currents the boundaries between the static low-resistance mode (SLR), the static high-resistance mode (SHR), and the region of thin-layer precession (PM) are similar to our previous results for Py/Cu/Co devices.[24] In addition, the Py/Cu/Py devices we have examined in this paper exhibit a critical current ($I_{c2}(H)$) beyond which the thick layer begins to precess (region T), as observed both from the onset of dips[28] in $dV/dI$ (Fig. 2(b)) and the appearance of narrow peaks in the microwave spectrum. The gray open squares denote the lines of dips in $dV/dI$ associated with jumps in the precession frequency of the thick layer.

The critical current for the onset of spin-transfer-driven dynamics for the thin magnetic layer, for perpendicular $H > 4\pi M_{thin}$, is predicted to have the form[18]

$$I_c(H) = \alpha_{thin}\gamma e S_{thin}(H - 4\pi M_{thin})/g(0), \quad (5)$$

where $\alpha$ is the Gilbert damping paramenter, $\gamma$ is the gyromagnetic ratio, $S_{thin}$ is the total spin of the thin layer, and $g(0)$ is an efficiency parameter associated with the amount of torque per unit current.[1] By a similar argument, the critical current for the onset of thick layer dynamics should be

$$I_{c2}(H) = \alpha_{thick}\gamma e S_{thick}(H - 4\pi M_{thick})/g(\pi). \quad (6)$$

The measured slopes of the critical currents are $d[I_c(H)]/dH = 0.168\pm0.001$ mA/T and $d[I_{c2}(H)]/dH = 0.912\pm0.052$ mA/T. We will compare the measured ratio of the slopes $\{d[I_{c2}(H)]/dH\}/\{d[I_c(H)]/dH\} = 5.4$ to the expectation from theory:

$$\frac{d[I_{c2}(H)]/dH}{d[I_c(H)]/dH} = \frac{S_{thick}}{S_{thin}} \frac{\alpha_{thick}}{\alpha_{thin}} \frac{g(0)}{g(\pi)}. \quad (7)$$

To evaluate this expression, we use $4\pi M_s = 7.8$ kOe for the thin layer and $4\pi M_s = 8.5$ kOe for the thick layer, based on SQUID measurements of test samples, and that the ratio of thicknesses for the two layers is 10. If the spin polarization of the current acting on each of the two layers is the same and equal to 30%, then using the formalism in ref. [1] we have $g(0) = 0.11$ and $g(\pi) = 0.37$. If we further assume that the Gilbert damping parameters for the two layers are similar, then the estimated ratio of the critical-current slopes is 2.9, almost a factor of 2 smaller than the measured value. The difference might be associated with a difference in the damping coefficients for the two layers, but this would require that the thicker layer experience more damping, a trend opposite to expectations within the theory of spin pumping.[40] We believe it is more likely that the spin-transfer efficiency parameters $g(0)$ and $g(\pi)$ are more nearly equal than predicted by ref. [1]. This supposition has been suggested by recent comparisons of the critical currents and switching speeds for P-to-AP and AP-to-P spin-transfer-driven switching.[39]

## COMPARISON WITH SINGLE-DOMAIN MAGNETIC SIMULATIONS

We have performed numerical simulations of the magnetic dynamics on our samples by integrating the Landau-Lifshitz-Gilbert equations of motion with the Slonczewski spin-transfer-torque term[1] applied to both magnetic layers. For simplicity, we assume the macrospin approximation -- that both magnetic layers can be treated as single domains without internal structure. However, we do not wish to claim that this is an accurate approximation. In fact, we have already described direct evidence for the excitation of non-uniform spin wave modes in the thick layer (modes B and B1). Nevertheless, it is interesting to analyze what features of the experimental phase diagram can be understood within the simplest model. A more accurate treatment would require a full micromagnetic analysis. We use the following parameters in our simulations: saturation magnetizations $4\pi M_{thick}$ = 8.5 kOe and $4\pi M_{thin}$ = 7.8 kOe (based on SQUID measurements of test samples), out-of-plane anisotropy $4\pi M_{eff}$ = 3.2 kOe and in-plane anisotropy $H_{an}$ = 2 kOe for the thick layer (based on fits to the precession frequencies), for the thin layer $4\pi M_{eff} \approx$ 4.5 kOe (based on fits to the precession frequencies) and $H_{an}$ = 0.2 kOe (determined from the switching field of thin layers in similar samples at 4.2 K), and a dipolar field acting on the thin layer $H_d$ = 0.5 kOe (determined from the room-temperature switching field of the superparamagnetic free layer). For the Gilbert damping parameter and the spin polarization of the current we use approximate bulk values: $\alpha$ = 0.01 and $P$ = 30%. These could be quite different in the confined geometry of the nanopillar devices. Nevertheless, we reach quite good qualitative agreement between the model and experiment data, as shown in Fig. 4(b) for an in-plane applied magnetic field and Fig. 6(b) for a perpendicular field.

The simulated phase diagram for in-plane $H$ gives a good account of the existence and relative positions of the different static and dynamical magnetic states that we observe experimentally. It describes well the transition between the P state and dynamical states at 0.6 kOe < $H$ < 4.4 kOe, the transition between the P state and the static high resistance state at large $H$, the existence of thick-layer precessional modes (the regions T), the position of the "clam-shell" precession regime (shaded) in which the precession frequency decreases with $I$, and the existence of the larger dynamical regime in which the frequency increases with $I$ (region D). The simulations indicate that region D corresponds to large-amplitude, out-of-plane, approximately circular precessional orbits for the thin layer. The maximum microwave power predicted by the simulations for this mode is 2.9 pW/(mA)$^2$ at 0.7 kOe, in reasonable agreement with the large powers measured experimentally (1.7 pW/(mA)$^2$ at 0.7 kOe). As a function of increasing $H$, the simulations predict that the microwave power should decrease (to 2.2 pW/(mA)$^2$ at 1.3 kOe), which is also in accord with the measurements described above in section IV.B. It is unclear why the circular out-of-plane precessional orbit was not present in past studies of devices containing Co thin layers, while it is present in the LLG simulations and for Py thin layers. We suspect that the dynamics of Py layers remain closer to single-domain than do the dynamics of Co layers, once the layer moment develops a significant component out-of-plane.

Despite the generally good agreement between the measurements and the single-domain simulations in Figs. 4(a) and 4(b), there are some differences. The simulation predicts a significant slope as a function of $I$ for the boundary between the circular out-of-plane precession (region D) and the static high-resistance state. Experimentally, the microwave signals become very small in this region, as the orbit of the out-of-plane precession shrinks to approach the static out-of-plane high resistance configuration, so that it is difficult to determine the precise phase boundary. At low $H$ and positive $I$ we also did not experimentally observe a clear boundary between the static AP state and the state of thick-layer precession T, although in this regime the signals associated with thick-layer precession might be small and the onset of the signal might be difficult to observe. The simulations predict a much more extensive region of thick-layer precession (region T) than we measure for magnetic fields above 5 kOe. However, the thresholds for the excitation of thick-layer precession in our simulation are very sensitive to the assumptions one makes about the degree of spin polarization for the current and the angular dependence of the spin torque[7] and the damping.[40] Additional analysis of spin-transfer effects at large currents might provide new insights to these issues.

The comparison between the simulation and the experimental results for $H$ applied perpendicular to the sample plane (Fig. 6) reveals good qualitative agreement similar to the case of in-plane field. All of the static and dynamical states seen experimentally are reproduced in the simulation, with the correct relative positions. One notable difference is that the single-domain simulations do not predict the jumps in frequency associated with the dips in $dV/dI$ within region T for the thick-layer precession.

## CONCLUSIONS

We have examined spin-transfer-driven magnetic excitations in (thin Py)/Cu/(thick Py) nanopillar devices in which both magnetic layers were etched through. The use of Py, with its relatively low magnetization and magnetic anisotropy relative to Co, and the fact that the thick Py layer in our device was not exchange-coupled to a continuous film, both reduce the current thresholds required to excite magnetic dynamics. We observe excitations of the thin layer, similar in most respects to previous studies, except that we observe one mode (circular out-of-plane precession, mode D) that is predicted by simulations but was not present in past measurements of Co/Cu/Co devices. We find in addition that dynamical modes involving the motion of the thick-layer moment can be excited at large values of current when the sign of the spin-transfer torque is such that it destabilizes the static orientation of the thick layer. Under our convention for the sign of the current, this destabilization occurs when the thin-layer moment is approximately antiparallel to the thick-layer moment and we apply positive current, or when the two moments are approximately parallel and we apply a negative current. These conditions are opposite to those observed to drive thin-layer dynamics. Dynamical modes involving the motion of the thick layer are distinguished by narrower peaks in the microwave spectrum than for the thin-layer dynamics, and also quite different frequencies. Generally, when the thick layer is excited by spin transfer torques, we observe multiple peaks in the microwave spectrum of resistance oscillations that are not harmonically related. This suggests that the thick layer is not precessing as a uniform single domain, but rather that spatially non-uniform spin-wave modes may also be excited. However, single-domain simulations of the magnetic dynamics can still explain most of the qualitative features of the dynamical phase diagrams as a function of current and magnetic field.

The microwave oscillations produced by the thick-layer dynamics can have a frequency that is much better defined than for the thin-layer modes. Even at room temperature, the thick layer modes can have $f/\Delta f > 1000$, where $\Delta f$ is the FWHM, with very little background. Furthermore, because these resistance oscillations occur at large current values, they can produce larger microwave signals than the thin-layer modes. Modes involving precession of the thick layer may therefore be the proper choice for use in applications such as spin-transfer-driven microwave sources and oscillators.

We acknowledge support from the NSF/NSEC program through the Cornell Center for Nanoscale Systems, from DARPA through Motorola, and from the Army Research Office. We also acknowledge NSF support through use of the Cornell Nanofabrication Facility/NNIN and the Cornell Center for Materials Research facilities.

# References

1. J. C. Slonczewski, J. Magn. Magn. Mater. **159**, L1 (1996).
2. L. Berger, Phys. Rev. B **54**, 9353 (1996).
3. J. C. Slonczewski, J. Magn. Magn. Mater. **195**, L261 (1999).
4. J. Z. Sun, Phys. Rev. B **62**, 570 (2000).
5.  A. Brataas, Yu. V. Nazarov, and G. E. W. Bauer, Phys. Rev. Lett. **84**, 2481 (2000); Eur. Phys. J. B **22**, 99 (2001).
6. Y. B. Bazaliy, B. A. Jones, and S. C. Zhang, J. Appl. Phys. **89**, 6793 (2001).
7. J. C. Slonczewski, J. Magn. Magn. Mater. **247**, 324 (2002).
8. M.D. Stiles and A. Zangwill, Phys. Rev. B. **66**, 014407 (2002).
9.  Z. Li and S. Zhang, Phys. Rev. B **68**, 024404 (2003).
10. M. Tsoi *et al*., Phys. Rev. Lett. **80**, 4281 (1998).
11. E. B. Myers *et al*., Science **285**, 867 (1999).
12. M. Tsoi *et al*., Nature **406**, 46 (2000).
13. W. H. Rippard *et al*., Phys. Rev. Lett. **92**, 027201 (2004).
14. J. A. Katine *et al*., Phys. Rev. Lett. **84**, 3149 (2000).
15. F. J. Albert *et al.* Appl. Phys. Lett. **77**, 1357 (2000).
16. J. Grollier *et al*., *Appl. Phys. Lett*. **78**, 3663 (2001).
17. J. E. Wegrowe *et al*., *Appl. Phys. Lett*. **80**, 3775 (2002).
18. B. Özyilmaz *et al*., Phys. Rev. Lett. **91**, 067203 (2003).
19. S. I. Kiselev *et al*., Nature **425**, 380 (2003).
20. S. Urazhdin, N. O. Birge, W. P. Pratt, and J. Bass, Phys. Rev. Lett. **91**, 146803 (2003).
21. R. H. Koch, J. A. Katine, and J. Z. Sun, *Phys. Rev. Lett*. **92**, 088302 (2004).
22.  M. Covington, A. Rebei, G. J. Parker, and M. A. Seigler, *Appl. Phys. Lett*. **8**, 3103 (2004).
23. M.R. Pufall *et al.*, *Phys. Rev. B* **69**, 214409 (2004).
24. S. I. Kiselev *et al., Phys. Rev. Lett*. **93**, 036601 (2004).
25. M. Tsoi, J. Z. Sun, and S. S. P. Parkin, *Phys. Rev. Lett*. **93**, 036602 (2004).
26. I. N. Krivorotov *et al. Phys. Rev. Lett*. **93**, 166603 (2004).
27. I. N. Krivorotov *et al*., *Science* **307**, 228 (2005).
28. B. Özyilmaz *et al*., cond-mat/0407210.
29. F. J. Albert, Ph. D. Thesis, Cornell University (2003).
30.  A. Brataas, Y. Tserkovnyak, and G. E. W. Bauer, cond-mat/0501672.
31.  Much narrower spectral peaks have been observed for Py thin layers in nanopillar samples in which exchange biasing was used to produce an offset angle between the thin-layer and thick-layer magnetic moments.  Precessional peaks as narrow as 10 MHz have been measured at 40 K in such samples.  See ref. [27].
32. C. Kittel, *Introduction to Solid State Physics*, **7**th edition (John Wiley & Sons, Inc, New York, 1996), p.505.
33.  J.A. Osborn, Phys. Rev. **67,** 351 (1945).
34.   J. C. Sankey (unpublished).

35. S. E. Russek, S. Kaka, W. H. Rippard, M. R. Pufall, and T. J. Silva, Phys. Rev. B **71**, 104425 (2005).
36. M. L. Polianski and P. W. Brouwer, Phys. Rev. Lett. **92**, 26602 (2004).
37. M. D. Stiles, J. Xiao, and A. Zangwill, Phys. Rev. B **69**, 054408 (2004).
38. C. Herring and C. Kittel, Phys. Rev. **81**, 869 (1951).
39. P. M. Braganca et al., preprint.
40. Y. Tserkovnyak *et al*., *Phys. Rev.* B **67**, 140404 (R) (2003).

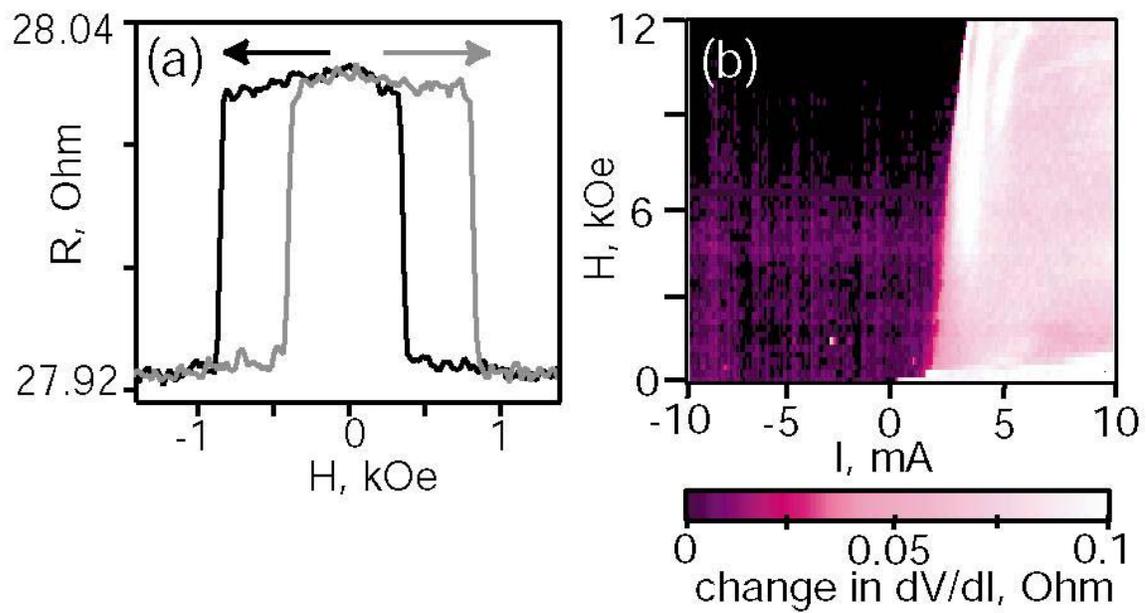

**Fig. 1** (**a**) Magnetoresistance near $I$=0 and (**b**) variations in $dV/dI$ as a function of $I$ for in-plane $H$. The resistance values listed in (a) include approximately 26 $\Omega$ of lead resistance.

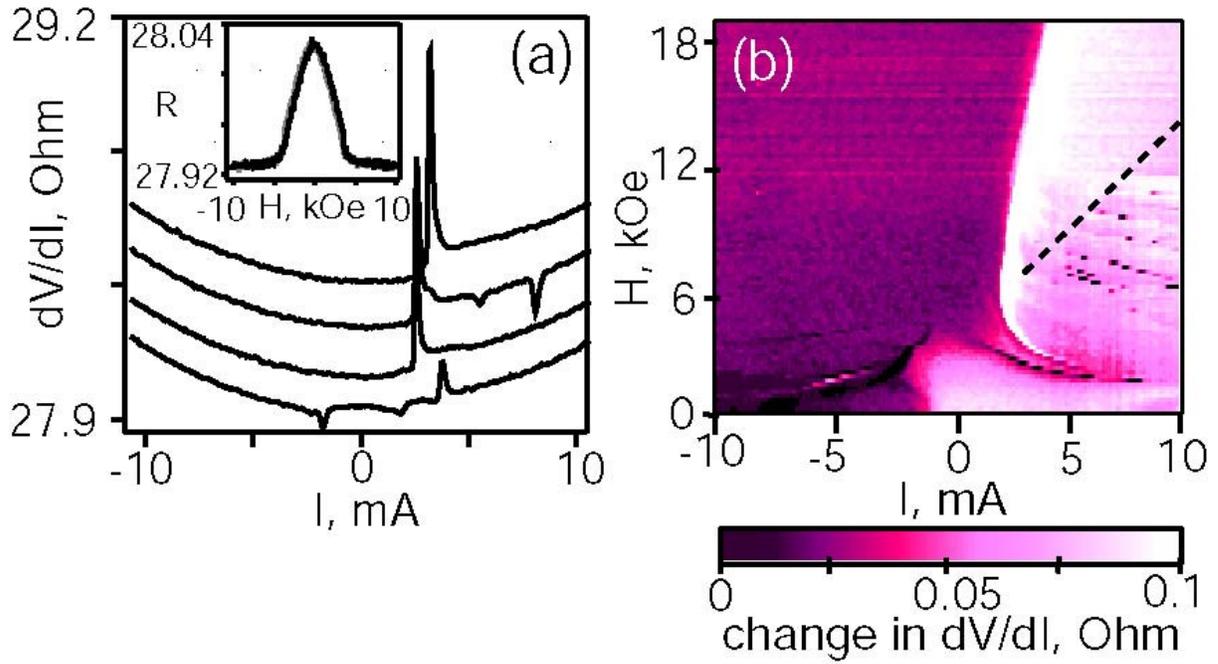

**Fig. 2 (a)** *dV/dI* at *H* = 3, 5, 7, and 10 kOe applied perpendicular to the sample plane. Curves are offset for clarity. Inset: Magnetoresistance near *I*=0. **(b)** Variations in *dV/dI* as a function of *I* for perpendicular *H*. The dashed line shows the position of the $I_{c2}$ line described in the text.

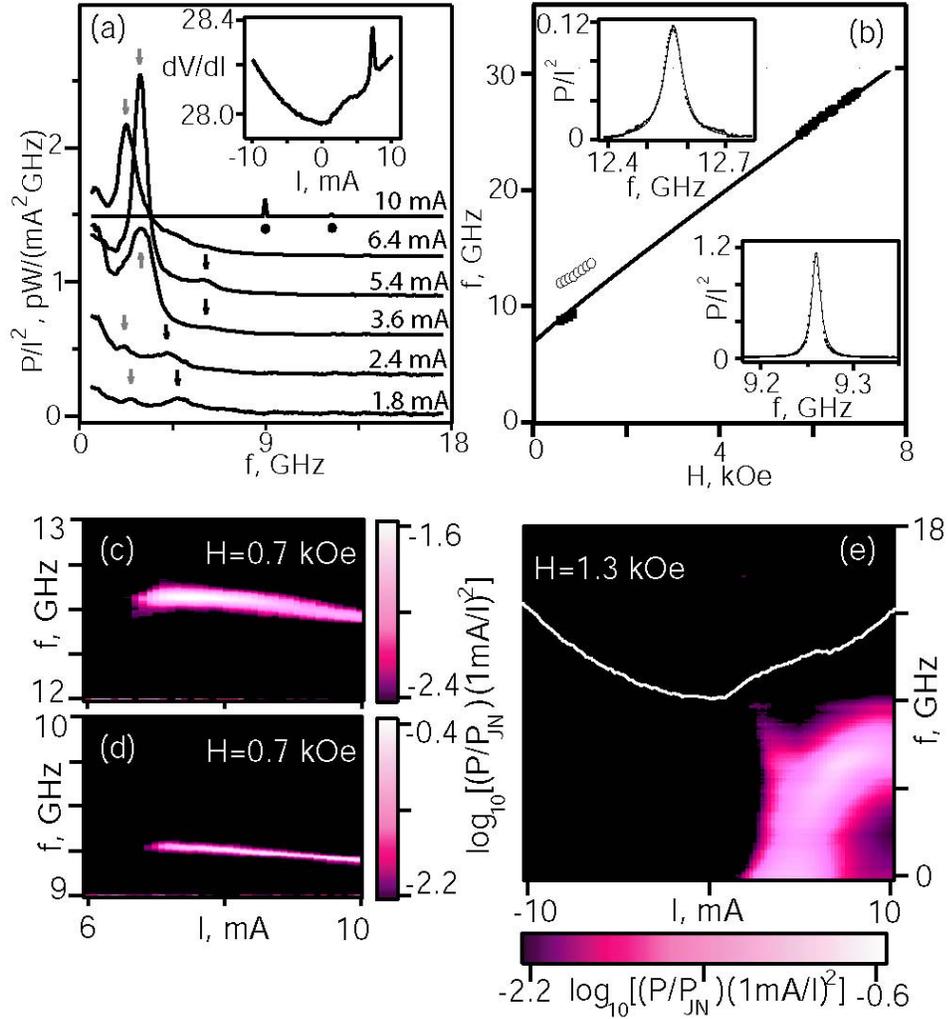

**Fig. 3 (a)** Microwave power density at in-plane $H = 0.7$ kOe along the magnetic easy axis at $I$ = 1.8, 2.4, 3.6, 5.4, 6.4 and 10 mA, from bottom to top. Curves are offset for clarity. Gray and black arrows show the position of the first and second harmonics of broad microwave signals. Dots under the $I = 10$ mA curve show the positions of two sharp microwave peaks. Inset: $dV/dI$ at $H = 0.7$ kOe. **(b)** Magnetic field dependence of the sharp microwave peaks at the values of $I$ for which they first appear. Open (closed) symbols describe the higher (lower) frequency mode. The solid line is a fit to Eq. (1). Insets: details of the sharp microwave peaks at $H = 0.7$ kOe and $I = 7$ mA. Microwave power density is measured in units of pW/mA$^2$GHz. **(c,d)** Current dependence of the microwave power density at $H = 0.7$ kOe for the sharp peaks. **(e)** Microwave power density at $H = 1.3$ kOe. The white line shows $dV/dI$.

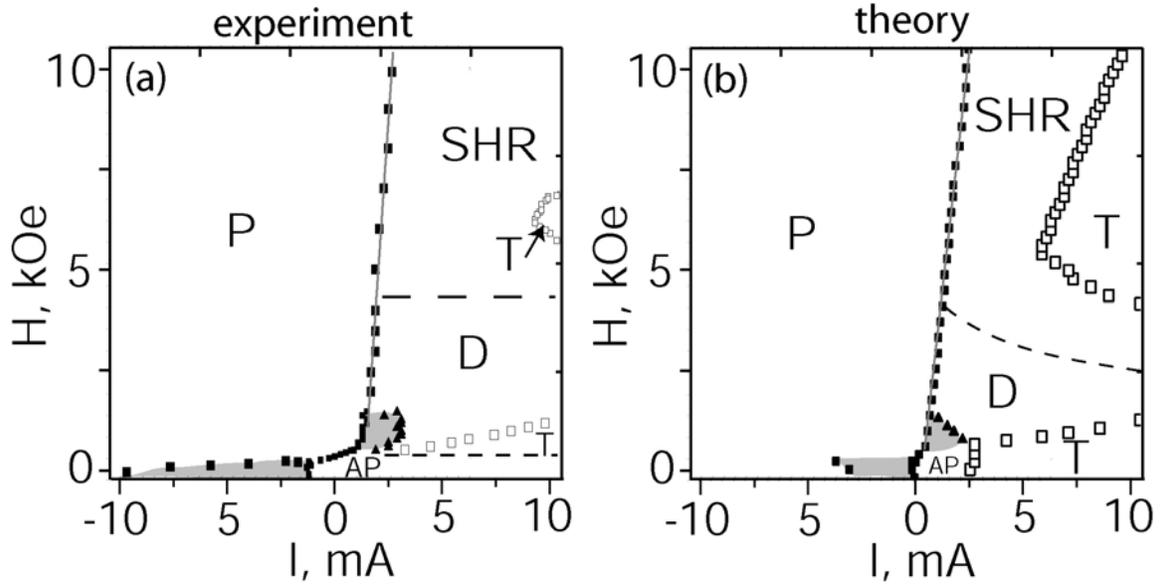

**Fig. 4 (a)** Experimental room-temperature dynamical phase diagram for in-plane magnetic field. P, AP, and SHR denote the parallel, antiparallel and static high resistance states. Gray areas mark regions of large angle "clam-shell" precession of the thin layer, where the frequency of oscillations decreases with increased |*I*|. Triangles show the transition from "clam-shell" precessional state to a regime of circular-out-of plane precession of the thin layer (marked by D). Open symbols show the onset of the sharp precessional signals that we associate with thick-layer dynamics (region T). **(b)** Theoretical room-temperature phase diagram obtained by numerical solution of the LLG equation with the parameters for the thin and thick layers described in the text.

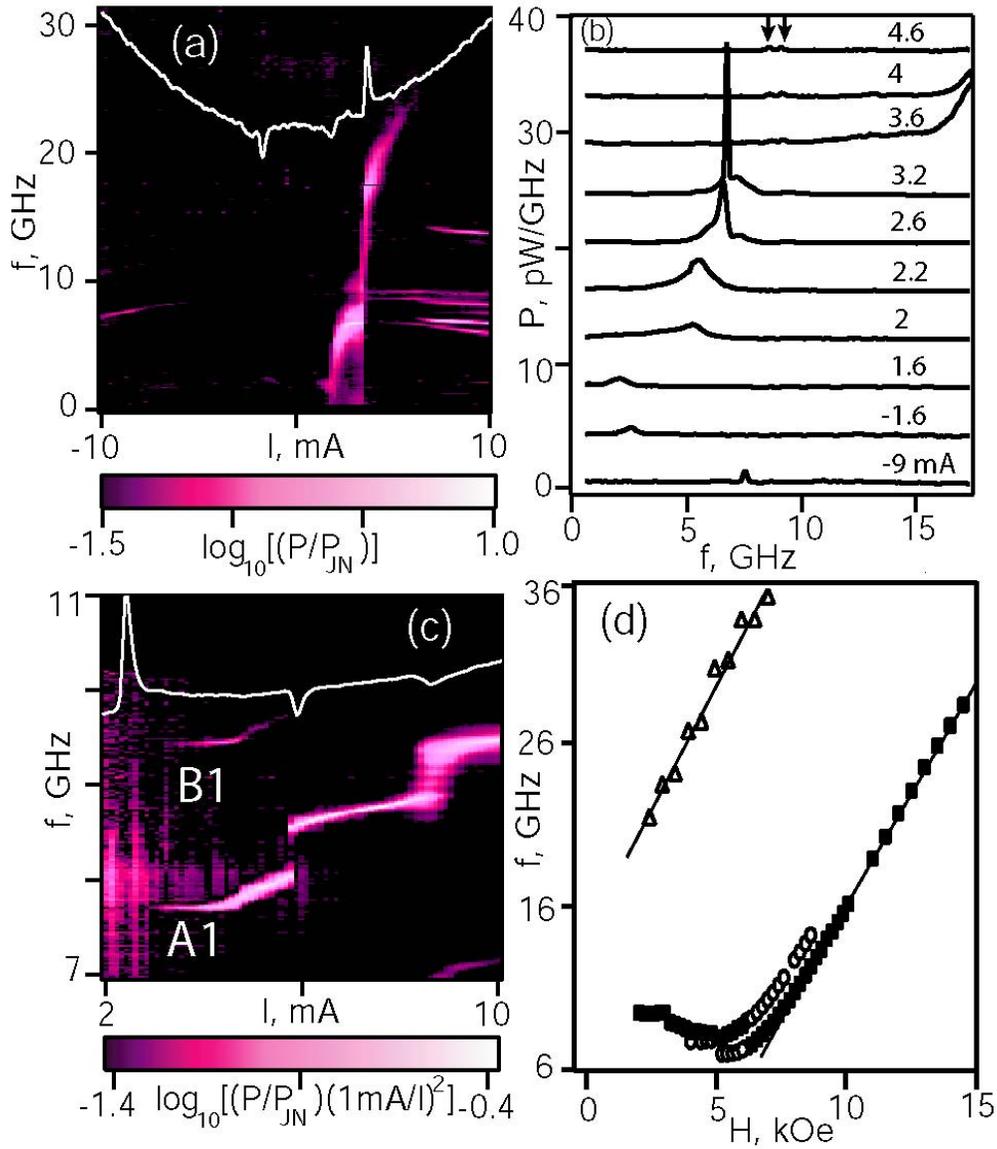

**Fig. 5 (a)** Microwave power density at perpendicular $H$ = 3 kOe. The white line shows $dV/dI$. **(b)** Individual traces of power density at perpendicular $H$ = 3 kOe measured at $I$ = -9.0, -1.6, 1.6, 2.0, 2.2, 2.6, 3.2, 3.6, 4.0, and 4.6 mA, from bottom to top. For display purposes, the curves are offset vertically and the amplitudes are multiplied by the factors 2, 4, 2, 1, 1, 0.5, 0.5, 2, 2, and 2, and were not corrected by $I^2$. Arrows point to the position of sharp modes appearing right after the frequency jump of the thin layer precession. **(c)** Microwave power density at $H$ = 7 kOe. The white line shows $dV/dI$. **(d)** Magnetic field dependence of microwave peaks. Open circles and filled squares show the position of sharp microwave peaks. The filled symbols denote the peak of the larger amplitude. The solid line is a fit to Eq. (2). Open triangles show the maximum thin-layer precessional frequency. The line associated with these data is a fit to Eq. (3).

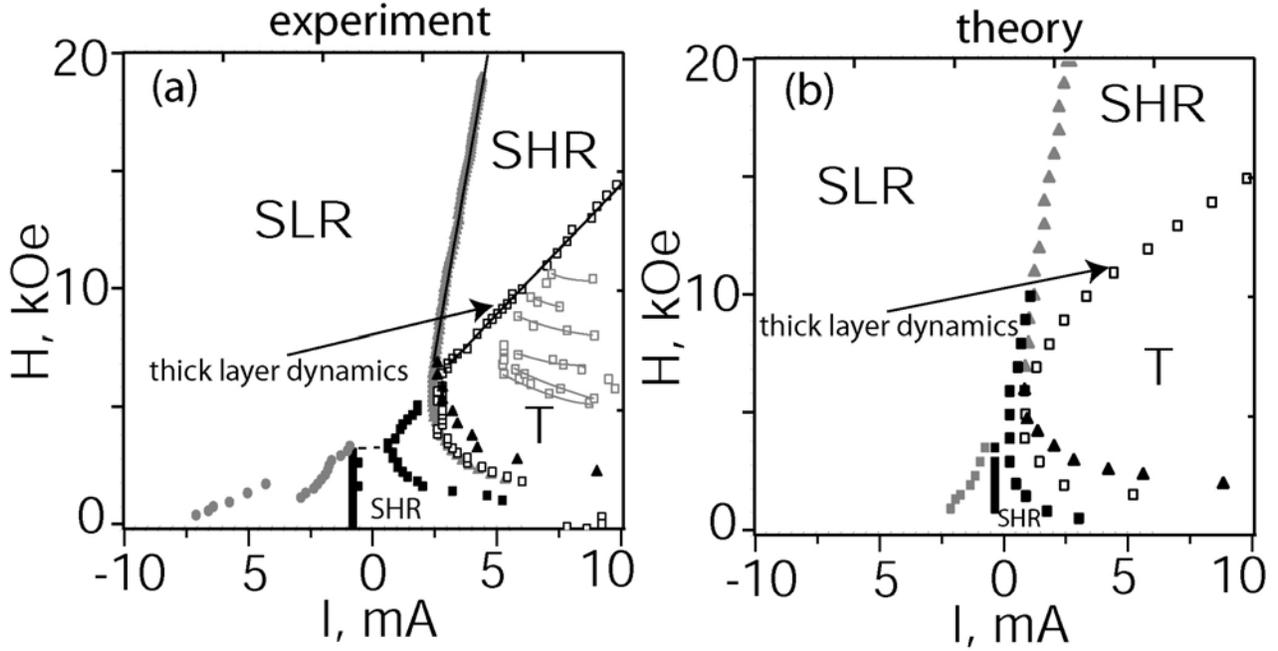

**Fig. 6 (a)** Experimental room-temperature dynamical phase diagram for perpendicular *H*. SHR and SLR mark static high and low resistance states. At negative *I*, gray symbols mark a transition between the SLR state and a dynamical regime of the thin permalloy layer, and closed black squares mark switching between the dynamical regime and the SHR state. At positive *I*, black squares show the transition between the SHR state and the dynamical state of thin layer. Gray triangles show the position of the peak in *dV/dI* which coincides with switching of the thin layer precession axis from parallel to antiparallel alignment with *H* at *H* < 5 kOe and with switching from the SLR state to the SHR state at *H* > 5 kOe. Closed triangles mark the turning-off of the thin layer precession. Open squares show the onset of sharp dynamical peaks corresponding to thick-layer precession (region T). Open gray symbols show the position of the dips in *dV/dI* corresponding to the steps in frequency for the thick layer precession. The gray lines are guides for the eye. Black lines show linear fits to the thresholds for dynamical instabilities of the thin and thick layers (Equations (5) and (6)). **(b)** Theoretical room-temperature phase diagram for perpendicular *H* obtained by numerical solution of the LLG equation with the parameters for the thin and thick layers described in the text.